\documentclass[prd,nofootinbib,twocolumn]{revtex4-1}

\usepackage{amssymb,amsthm}
\usepackage{epsfig}

\newcommand\be{\begin{eqnarray}}
\newcommand\ee{\end{eqnarray}}

\newcommand{\C}{{\mathbb C}}
\newcommand{\im}{{\rm i}}

\begin{document}

\title{New Action Principle for General Relativity}

\author{Kirill Krasnov}
\affiliation{School of Mathematical Sciences, University of
Nottingham, Nottingham, NG7 2RD, UK.}

\date{March 2011}

\begin{abstract} General Relativity can be reformulated as a diffeomorphism invariant ${\rm SU}(2)$ gauge theory. A new action principle for this "pure connection" formulation of GR is described. 
\end{abstract}

\maketitle

In \cite{Plebanski:1977zz} Plebanski has shown that instead of the spacetime metric the dynamical field of General Relativity (GR) can be taken to be a collection (triple) of two-forms satisfying a certain algebraic equation. This idea was taken further in \cite{Capovilla:1989ac}, \cite{Capovilla:1991kx}, where it was suggested that the two-form field (as well as the Lagrange multiplier field of the Plebanski formulation) can be integrated out to obtain a "pure connection formulation" of GR. An action principle realizing this idea in the case $\Lambda=0$ of zero cosmological constant GR was described in \cite{Capovilla:1989ac}. The case $\Lambda\not=0$ proved to be more difficult, and only a rather complicated (and erroneous, see \cite{CDJ-erratum}) action was given in \cite{Capovilla:1991kx}. The purpose of this letter is to point out that an elegant and simple "pure connection" action principle for GR encompassing both $\Lambda\not=0$ and $\Lambda=0$ cases is possible. 

We start by describing the new variational principle, and then prove that solutions of the arising Euler-Lagrange equations are in one-to-one correspondence with solutions of Einstein equations. As in \cite{Capovilla:1989ac}, \cite{Capovilla:1991kx}, the main dynamical field of our formulation is an ${\rm SO}(3)$ connection field $A^i, i=1,2,3$ over the spacetime manifold. We take $A^i$ to be a dimensionful quantity with dimensions of $A\sim 1/L$, where $L$ is a unit of length. In contrast to \cite{Capovilla:1989ac}, \cite{Capovilla:1991kx}, our action will be a functional of only the connection; no additional auxiliary field will be necessary. Given a connection $A^i$, its curvature is given by $F^i=dA^i + (1/2)\epsilon^{ijk} A^j\wedge A^k$. Here the form notation is used, and $\wedge$ denotes the wedge product of forms. As in \cite{Capovilla:1989ac}, \cite{Capovilla:1991kx} we consider the 4-form $F^i\wedge F^j$, which is valued in the second (symmetric) power of the Lie algebra ${\mathfrak su}(2)$. Using the density weight one anti-symmetric tensor $\tilde{\epsilon}^{\mu\nu\rho\sigma}$, which does not need a metric for its definition (here $\mu,\nu,\ldots$ are the spacetime indices), we can convert the 4-form $F^i\wedge F^j$ into a density weight one symmetric $3\times 3$ matrix 
\be\label{X}
\tilde{X}^{ij}:= \frac{1}{4} \tilde{\epsilon}^{\mu\nu\rho\sigma} F_{\mu\nu}^i F_{\rho\sigma}^j,
\ee
so that $F^i\wedge F^j = \tilde{X}^{ij} d^4x$. We note that $\tilde{X}^{ij}$ has dimensions of $1/L^4$. Now consider an arbitrary homogeneous of degree one, gauge invariant function $f: {\rm Mat}(3\times 3)\to \C$, i.e., a function satisfying $f(\alpha \tilde{X}) = \alpha f(\tilde{X})$ as well as $f(O \tilde{X} O^T) = f(\tilde{X}), O\in {\rm SO}(3)$.  Then $f(\tilde{X}^{ij})$ is a density weight one, and can be integrated over the spacetime to produce an action. We refer the reader to e.g.  \cite{Krasnov:2011up} for more details on this construction of diffeomorphism invariant actions. We also note that the sketched construction of actions is somewhat similar to that described in \cite{Hitchin:2001rw} in the context of stable differential forms. 

The simplest possible diffeomorphism invariant gauge theory action corresponds to $f(\tilde{X})={\rm Tr}(\tilde{X})$. This, however, gives a topological theory without any interesting dynamics. As we shall now see, general relativity (with $\Lambda\not=0$) arises for a certain other choice of $f$. To describe it, let us recall the notion of the matrix square root. For a $3\times 3$ symmetric matrix such as (\ref{X}), a square root $(\sqrt{\tilde{X}})^{ij}$ is a symmetric matrix such that $(\sqrt{\tilde{X}})^{ij} (\sqrt{\tilde{X}})^{jk}=\tilde{X}^{ik}$. Explicitly, if the matrix $\tilde{X}^{ij}$ is diagonalized by an orthogonal transformation $O\in{\rm SO}(3)$, i.e., $\tilde{X}=O D O^T, D={\rm diag}(\tilde{\lambda}_1,\tilde{\lambda}_2,\tilde{\lambda}_3)$, then $(\sqrt{\tilde{X}})=O \sqrt{D} O^T, \sqrt{D}={\rm diag}(\sqrt{\tilde{\lambda}_1},\sqrt{\tilde{\lambda}_2},\sqrt{\tilde{\lambda}_3})$. This involves a choice of the branch of the square root function. For our purposes any of the two branches can be taken; we shall see that the action is independent of this choice. Indeed, consider the function 
\be\label{f}
f(\tilde{X}) := \frac{1}{16\pi G\Lambda} \left( {\rm Tr}\sqrt{\tilde{X}} \right)^2.
\ee
Here $G,\Lambda$ are the Newton's and cosmological constants respectively. Note that because of the second power present here the function (\ref{f}) is independent of which branch of the square root is used. The function (\ref{f}) is homogeneous of degree one and gauge invariant. Thus, it satisfies all the requirements discussed above, and so $f(\tilde{X})$ can be integrated over the spacetime to produce an action. We note that in the units $c=1$ used in this letter the quantity $1/(G\Lambda)$ has dimensions of $\hbar$. Thus, when (\ref{f}) is integrated over the manifold the result will have dimensions of $\hbar$, as is appropriate for the action. 

Having in mind the construction just described, we can write the following diffeomorphism invariant functional of the connection:
\be\label{action}
S_{\rm GR}[A] = \frac{1}{16\pi\im\, G\Lambda} \int  \left( {\rm Tr}\sqrt{F^i\wedge F^j} \right)^2.
\ee
This is the action principle that is the subject of this letter. Here $\im=\sqrt{-1}$ is the imaginary unit. The fact that there is a multiple of the imaginary unit in front of the action has to do with (salient up to now) fact that in the physically relevant case of spacetimes with metrics of Lorentzian signature the ${\rm SO}(3)$ connection field in (\ref{action}) is complex-valued. As in \cite{Plebanski:1977zz}, \cite{Capovilla:1991kx}, the action is supplemented with the reality conditions:
\be\label{reality}
F^i \wedge (F^j)^* = 0, \qquad {\rm Re} (F^i\wedge F^i) = 0,
\ee
which guarantee that real Lorentzian signature metrics arise. The first of the reality conditions in (\ref{reality}) says that the curvature must be wedge-orthogonal to its complex conjugate. The second condition says that the 4-form $F^i\wedge F^i$ is purely imaginary. This explains why a factor of $(1/\im)$ is needed in (\ref{action}). Note that there are 10 reality conditions in (\ref{reality}), exactly the number that is needed to require 10 metric components to be real. We also note the changes that are necessary to get an action that describes the Riemannian signature sector of general relativity. In this case the connection field is real ${\rm SO}(3)$-valued, no reality conditions (\ref{reality}) is necessary, and there is no factor of $1/\im$ in front of the action. In both the Lorentzian and Riemannian sectors we shall consider the variational principle (\ref{action}) for $A^i$ varying in the subspace ${\rm det}(\tilde{X})\not=0$, which guarantees that the spacetime metrics described by our connection field are non-degenerate. 

To prove that (\ref{action}) is indeed an action for general relativity in disguise, we need to find the corresponding Euler-Lagrange equations. The matrix of partial derivatives of the function (\ref{f}) with respect to the components of the matrix $\tilde{X}^{ij}$ is given by:
\be\label{partial}
\frac{\partial f}{\partial \tilde{X}^{ij}} = \frac{1}{16\pi G\Lambda} \left({\rm Tr}\sqrt{\tilde{X}}\right) \left( \sqrt{\tilde{X}}\,{}^{-1}\right)^{ij}.
\ee
Note that the inverse of $\tilde{X}^{ij}$ exists as is guaranteed by our restriction ${\rm det}(\tilde{X})\not=0$. Note also that, as is appropriate for a function of degree of homogeneity one, we have:
\be
\frac{\partial f}{\partial \tilde{X}^{ij}} \tilde{X}^{ij} = f(X).
\ee
If we now define:
\be\label{B}
B^i:= \frac{\partial f}{\partial \tilde{X}^{ij}} F^j,
\ee
where the matrix of partial derivatives is evaluated at (\ref{X}), then the Euler-Lagrange equations for (\ref{action}) read:
\be\label{EL}
D_A B^i = 0.
\ee
Here $D_A$ is the covariant derivative with respect to the connection $A^i$. We note that (\ref{EL}) is a set of $3\times 4$ second-order differential equations for the $3\times 4$ components of the connection $A^i$. 

We now show that (\ref{EL}), together with the definition (\ref{B}) of the two-form field $B^i$ are equivalent to the field equations of Plebanski formulation of GR \cite{Plebanski:1977zz}. To this end we note that the two-form field (\ref{B}) satisfies a set of algebraic equations. Indeed, using the definition (\ref{X}) we have:
\be\nonumber
\frac{1}{4} \tilde{\epsilon}^{\mu\nu\rho\sigma} B_{\mu\nu}^i B_{\rho\sigma}^j = \frac{\partial f}{\partial \tilde{X}^{ik}} \frac{\partial f}{\partial \tilde{X}^{jl}} \tilde{X}^{kl}=\left( \frac{{\rm Tr}\sqrt{\tilde{X}}}{16\pi G\Lambda} \right)^2 \delta^{ij}.
\ee
Thus, the two-form field (\ref{B}) satisfies:
\be\label{metricity}
B^i\wedge B^j \sim \delta^{ij},
\ee
which is the basic equation of Plebanski's formulation of GR \cite{Plebanski:1977zz}. The Einstein equations then arise as follows. Given a triple of two-forms $B^i$ satisfying (\ref{metricity}) there is a canonically defined spacetime metric (determined by the condition that $B^i$ are self-dual, and that a multiple of $B^i\wedge B^i$ is the volume form). This metric is real Lorentzian in view of (\ref{reality}). When ${\rm det}(\tilde{X})\not=0$ the equations (\ref{EL}) can be solved for the connection $A^i$ and the equation (\ref{metricity}) implies that the resulting ${\rm SO}(3,\C)$ connection is the self-dual part of the metric-compatible spin connection. The equation (\ref{B}) rewritten as
\be
F^i = \left( \frac{\partial f}{\partial \tilde{X}^{ij}} \right)^{-1} B^j
\ee
then implies that the curvature of the self-dual part of the spin connection is self-dual as a two-form, which is equivalent to the Einstein condition $R_{\mu\nu}\sim g_{\mu\nu}$. For more details on Plebanski formulation in the notations close to those of this letter the reader is referred to \cite{Krasnov:2009pu}. We have thus shown that the field equations following from (\ref{action}) are equivalent to those of the Plebanski formulation \cite{Plebanski:1977zz}. Thus, any solution of Einstein's theory gives rise to a solution of the theory (\ref{action}), and any (non-degenerate ${\rm det}(\tilde{X})\not=0$) solution of field equations of (\ref{action}) gives rise to an Einstein metric. 

We note that the logic of the above proof of equivalence to GR could be reversed, and one could {\it derive} (\ref{f}) as the only function of the matrix $\tilde{X}^{ij}$ of curvature wedge products such that the corresponding action produces Plebanski field equations. Indeed, it is clear that the key point about the particular choice (\ref{f}) is that it leads to (\ref{metricity}). This is the case for
\be
\frac{\partial f}{\partial \tilde{X}^{ij}} \sim \left({\rm Tr}\sqrt{\tilde{X}}\right) \left(\sqrt{\tilde{X}}\,{}^{-1}\right)^{ij},
\ee
where we need the trace prefactor in order to guarantee that $\partial f/\partial \tilde{X}$ is of degree of homogeneity zero. This then integrates to (\ref{f}). 

So far we have discussed the case of GR with non-zero cosmological constant. Indeed, in the limit $\Lambda\to 0$ the action (\ref{action}) is singular. However, in this limit the (exponential of the) action present in the quantum mechanical path integral of the theory can be viewed as a delta-function imposing the constraint:
\be
{\rm Tr}\sqrt{F^i\wedge F^j}=0.
\ee
This is the same equation as was found in \cite{Capovilla:1989ac} by rewriting the $\Lambda=0$ general relativity in the pure connection formalism. Indeed, the condition that the trace of the square root of a matrix is equal to zero can be rewritten as an equation on the matrix itself. For this we, as in \cite{Capovilla:1991kx}, denote $Y^{ij}\sim \sqrt{F^i\wedge F^j}$, and write down the characteristic equation for $Y$:
\be\nonumber
Y^3 - {\rm Tr}(Y) Y^2 + \frac{1}{2}\left(({\rm Tr}(Y))^2- {\rm Tr}(Y^2) \right) Y = {\rm det}(Y).
\ee
Now, multiplying by $Y$, taking the trace and using ${\rm Tr}(Y)=0$ we get:
\be
{\rm Tr}(Y^4) -\frac{1}{2}( {\rm Tr}(Y^2))^2=0.
\ee
Rewriting this in terms of $X=Y^2, X^{ij}\sim F^i\wedge F^j$ we get:
\be
{\rm Tr}(X^2)-\frac{1}{2} ({\rm Tr}(X))^2 = 0,
\ee
which is just the condition found in \cite{Capovilla:1989ac}, \cite{Capovilla:1991kx}. Thus, in the sense described, our action (\ref{action}) encompasses both $\Lambda\not=0$ and $\Lambda=0$ cases. 

Finally, we present an alternative derivation of the action (\ref{action}) directly from the Plebanski formulation of GR, via the same procedure as is followed in  \cite{Capovilla:1991kx}.  The Plebanski action for GR with a cosmological constant $\Lambda$ is a functional of the connection $A^i$, an ${\mathfrak su}(2)$-valued two-form $B^i$, as well as a field of Lagrange multipliers $\Psi^{ij}$. It is given by
\be\nonumber
S_{\rm Pleb}=\frac{1}{8\pi \im G}\int \left[B^i\wedge F^i - \frac{1}{2} \left( \Psi^{ij}+ \frac{\Lambda}{3}\delta^{ij}\right) B^i\wedge B^j\right].
\ee
More details on this formulation can be found in e.g. \cite{Krasnov:2009pu}. Integrating out the two-form field one gets the following action
\be
S[A,\Psi] = \frac{1}{16\pi\im G} \int \left( \Psi^{ij}+ \frac{\Lambda}{3}\delta^{ij}\right)^{-1} F^i\wedge F^j,
\ee
where it is assumed that the matrix $\left( \Psi^{ij}+ (\Lambda/3)\delta^{ij}\right)$ is invertible. It is now convenient to rescale the Lagrange multipliers field and write the action as
\be
S[A,\tilde{\Psi}] = \frac{1}{\im} \int \left(\tilde{\Psi}^{ij} + \alpha \delta^{ij} \right)^{-1} F^i\wedge F^j,
\ee
where
\be
\alpha:= \frac{16\pi G\Lambda}{3},
\ee
in units $\hbar=c=1$, is a dimensionless quantity. Note that $\alpha\sim M_\Lambda^2/M_p^2$ and so, for the observed value of the cosmological constant, is of the order $\alpha\sim 10^{-120}$. 

In the final step we integrate out the Lagrange multiplier field $\tilde{\Psi}^{ij}$. Let us drop the tilde on the symbol for brevity. We can the rewrite the above action as
\be
S[A,\Psi]=\frac{1}{\im} \int ({\rm vol}) {\rm Tr}\left((\Psi +\alpha {\rm Id})^{-1} X\right),
\ee
where we have introduced $F^i\wedge F^j = ({\rm vol}) X^{ij}$, and $({\rm vol})$ is an arbitrary auxiliary 4-form on our manifold. To integrate out the matrix $\Psi$ we have to solve the field equations for it, and then substitute the result back into the action. Assuming that the solution for $\Psi$ can be written as a matrix function of $X$, we conclude that $\Psi$ will be diagonal if $X$ is. Thus, we can simplify the problem of finding $\Psi$ by using an ${\rm SO}(3)$ rotation to go to a basis in which $X$ is diagonal. This is always possible at least locally. We then look for a solution in which $\Psi$ is also diagonal. Denoting by $\lambda_1,\lambda_2,\lambda_3$ the eigenvalues of $X^{ij}$, and by $a,b,-(a+b)$ the components of the diagonal matrix $\Psi$, we get the following action functional to consider
\be\label{app-1}
F[a,b,\lambda]= \frac{\lambda_1}{\alpha+a}+\frac{\lambda_2}{\alpha+b} + \frac{\lambda_3}{\alpha-(a+b)}.
\ee
We now have to vary this with respect to $a,b$ and substitute the solution back to obtain the defining function as a function of $\lambda_i$. Assuming that no of the denominators in (\ref{app-1}) are zero we get the following two equations
\be
(\alpha+a)^2 \lambda_3 = (\alpha-(a+b))^2\lambda_1, \\ \nonumber (\alpha+b)^2 \lambda_3 = (\alpha-(a+b))^2\lambda_2.
\ee
Taking the (positive branch of the) square root and adding the results we get $a+b$, which is most conveniently written as 
\be
\alpha-(a+b)=3\alpha \frac{\sqrt{\lambda_3}}{\sqrt{\lambda_1}+\sqrt{\lambda_2}+\sqrt{\lambda_3}}.
\ee
The other two combinations that appear in (\ref{app-1}) are given by similar expressions:
\be
\alpha+a = 3\alpha \frac{\sqrt{\lambda_1}}{\sqrt{\lambda_1}+\sqrt{\lambda_2}+\sqrt{\lambda_3}}, \\ \nonumber \alpha+b = 3\alpha \frac{\sqrt{\lambda_2}}{\sqrt{\lambda_1}+\sqrt{\lambda_2}+\sqrt{\lambda_3}}.
\ee
It is now clear that the sought function of the matrix $X$ is given by
\be\nonumber
f_{GR}(\lambda) = \frac{1}{3\alpha} \left( \sqrt{\lambda_1}+\sqrt{\lambda_2}+\sqrt{\lambda_3}\right)^2 = \frac{1}{3\alpha}\left({\rm Tr}\sqrt{X}\right)^2.
\ee
Integrated over the spacetime manifold this is just our action (\ref{action}). This concludes our proof of the classical equivalence of General Relativity with a non-zero cosmological constant and the theory of connections (\ref{action}).  

A number of remarks is in order. First, the formulation (\ref{action}) can be used as the starting point for a new type of the gravitational perturbation theory. Here one expands the action around the constant curvature background, and the usual linearized GR solutions (gravitons) can be seen to appear \cite{Krasnov:2011up}. It would be very interesting to develop this line of thought further and compute the graviton scattering amplitudes as well as loop corrections using this formalism. Work on these issues is in progress. 

Another remark is that the usual metric based GR and the theory of connections (\ref{action}) may well be different as quantum theories. Indeed, the transformation from the set of metric variables of GR to the set of connection components in (\ref{action}) is highly non-trivial, and so the natural path integral measures for the two theories do not have to coincide. It is thus not impossible that quantum mechanical calculations based on (\ref{action}) will produce quantitatively different results from those in perturbative GR. It will then be a matter of choice to select the "correct" quantum theory. Note, however, that the theory (\ref{action}) expanded around a constant curvature background is as non-renormalizable as GR (in that the coupling constant has a negative mass dimension), see \cite{Krasnov:2011up}.

Another (potentially important) point about the formulation (\ref{action}) is that it immediately allows for a very large class of generalisations. Indeed, we have seen that the construction of the action goes through for any homogeneous order one and gauge invariant function $f(\tilde{X})$. The function in (\ref{action}) is special because it guarantees (\ref{metricity}), but other functions can be considered. A large class of diffeomorphism invariant ${\rm SU}(2)$ gauge theories is then possible, see e.g. \cite{Krasnov:2011up} and \cite{Bengtsson:1990qg} for earlier work. A very interesting feature of all these new theories is that they describe just two propagating degrees of freedom, see \cite{Krasnov:2007cq}, exactly like general relativity. These theories could be of importance for understanding the ultra-violet behaviour of gravity, see \cite{Krasnov:2011up} for more details on such potential applications. 

Let us also comment on the fact that for the currently accepted value of the cosmological constant the coefficient $1/G\Lambda$ in front of the action (\ref{action}) is so exceedingly large (approximately $10^{120}$ if measured in the units of $\hbar$). Thus, in the language of our "pure connection" formulation the famous cosmological constant problem becomes a question of why the dimensionless parameter that multiplies the gravitational action is so enormous. It could be that the answer is given by an appropriate renormalisation group flow, see \cite{Krasnov:2011up} for more discussion. 

Note also that it is a very interesting point about (\ref{action}) that it requires a non-zero $\Lambda$. Given that there is now a strong observational evidence for a non-zero cosmological constant, this seems to be a step in the right direction as compared to the usual metric based GR whose action principle works equally well with or without $\Lambda$. 

Apart from possible applications in quantum gravity, the new formulation (\ref{action}) may prove instrumental in the classical domain. One promising direction appears to be to questions about the moduli spaces of Einstein metrics on 4-manifolds, see e.g. \cite{Besse}, Chapter 12. The point is that the linearization of the new functional (\ref{action}) behaves differently with respect to diffeomorphisms than the linearization of the Einstein-Hilbert functional. Indeed, at least for the linearization around the constant curvature background  \cite{Krasnov:2011up} one finds the linearized action to be simply indepedent of certain components of the connection (those which can be changed by an action of a diffeomorphism). This is completely different from the case of the Einstein-Hilbert functional, where diffeomorphisms need to be gauge-fixed in a rather non-trivial fashion. Optimistically, this different behaviour may make some open rigidity questions about Einstein metrics easier to tackle. 

Apart from the above positive features, the new formulation (\ref{action}) has some difficulties that must be mentioned. The first and foremost is that for applications in physics one needs to know how all other matter couples to gravity. It is not easy to describe this once gravity has been reformulated as a theory of connections. Fortunately, a simple way to couple the usual Yang-Mills gauge fields exists, see \cite{TorresGomez:2009gs}, \cite{TorresGomez:2010cd}, and also \cite{Chakraborty:1994vx} for earlier work. The main idea here is that enlarging the gauge group appropriately and expanding the theory around the constant curvature background in the gravitational sector, one finds the usual Yang-Mills action functional as describing the low energy physics of the non-gravitational gauge fields. It is much more difficult to couple to (\ref{action}) fermionic matter, and it is clear that new ideas will be required here. Work on this is in progress. 

Another difficulty with the formulation (\ref{action}), as well as with the original Plebanski formulation \cite{Plebanski:1977zz}, is that the connection field is required to be complex-valued (if one is to reproduce the Lorentzian signature sector of GR). Then reality conditions (\ref{reality}) need to be imposed, so that the action (\ref{action}) is varied among the gauge fields satisfying (\ref{reality}). For the classical theory this is not much of a problem, but if one wants to base on (\ref{action}) a quantum mechanical treatment, one has to take into account (\ref{reality}) in the path integral, which is a difficult task. One possible way around this problem could be to resort to the analytic continuation to the Riemannian signature metrics, where the reality is trivial to impose. However, it is not at all clear if there is a consistent way to do this in the quantum theory. More work on these issues is required. 

In spite of some questions remaining open, we feel that the action principle (\ref{action}) has a potential to become a starting point for new developments in both classical and quantum General Relativity. 

\section*{Acknowledgments} The author would like to thank Ingemar Bengtsson for useful comments and for pointing out the reference \cite{CDJ-erratum}.

\end{document}